\documentclass[letterpaper, amsfonts, amssymb, amsmath, reprint, showkeys, nofootinbib, twoside,superscriptaddress]{revtex4-1}
\usepackage[english]{babel}
\usepackage[utf8]{inputenc}
\usepackage{url}
\usepackage[pdftex]{hyperref} 
\usepackage{graphicx}

\newcommand{\papertitle}{Deterministic coupling of ultracold atomic lattice to a suspended photonic waveguide}

\newcommand{\linkcolor}{magenta}%
\hypersetup{colorlinks=true, %
  linkcolor=\linkcolor, %
  citecolor=\linkcolor, %
  filecolor=\linkcolor, %
  urlcolor=\linkcolor, %
  pdftitle=\papertitle,
  pdfauthor={J.-B. Béguin}, %
  pdfsubject={}, %
  pdfkeywords={} %
}
\usepackage{physics}
\usepackage{siunitx}
\usepackage{caption,subcaption}
\usepackage{ragged2e}
\DeclareCaptionJustification{justified}{\justifying}
\begin{document}
\title{\papertitle}

\author{J. T. Hansen}    
    \affiliation{Niels Bohr Institute, University of Copenhagen, Blegdamsvej 17, 2100 Copenhagen, Denmark}

\author{F. Gargiulo}    
    \affiliation{Niels Bohr Institute, University of Copenhagen, Blegdamsvej 17, 2100 Copenhagen, Denmark}
    
\author{J. B. Mathiassen}
    \affiliation{Niels Bohr Institute, University of Copenhagen, Blegdamsvej 17, 2100 Copenhagen, Denmark}
    
\author{J. H. M\"uller}
    \affiliation{Niels Bohr Institute, University of Copenhagen, Blegdamsvej 17, 2100 Copenhagen, Denmark}
    
\author{E. S. Polzik}
    \affiliation{Niels Bohr Institute, University of Copenhagen, Blegdamsvej 17, 2100 Copenhagen, Denmark}
    
\author{J.-B. B\'eguin}
    \email[Correspondence email address: ]{jbeguin@nbi.ku.dk}
    \affiliation{Niels Bohr Institute, University of Copenhagen, Blegdamsvej 17, 2100 Copenhagen, Denmark}

\date{\today}

\begin{abstract}
The deterministic control of light-matter interactions at the level of single particles and on subwavelength scales is central to quantum optics and hybrid integrated quantum technologies. However, combining cold atom research with nanophotonic devices in a fully controllable platform remains a major experimental challenge. Here, we demonstrate the deterministic coupling of an ultracold atomic lattice to light propagating in suspended on-chip photonic circuits. These capabilities open avenues to address scalability challenges in neutral-atom quantum computers and simulators, enabling fast optical readout, efficient and subwavelength non-diffracting  interaction zones, and genuine compatibility with integrated solid-state photon sources, detectors, and stop-band modulators. Beyond controllable quantum matter, the platform also enables in-situ imaging of evanescent fields of light and nanoscale structures, including prospects for three-dimensional scanning microscopy with non-invasive single-atom probes for quantum sensing applications.
\end{abstract}


\maketitle

\section{Introduction}

Recent worldwide advances in controlling individual cold atoms with optical tweezer lattices \cite{GrossBlochQuantumSimulation, VuleticLukinErrorDetected, Su2025, BrowaeysInteractingBosonsRydberg} are opening up an exciting new frontier in quantum physics. Beyond quantum simulation and computing, new phases of quantum matter \cite{Impertro2025} and new types of defect-free quantum dielectrics \cite{ana2019} could be assembled one atom at a time.

A promising and complementary approach to free-space trapped atoms and Rydberg atom-atom interactions \cite{Saffman2010} for quantum information science is to combine neutral atoms with nanophotonic circuits that can efficiently guide and control light. By coupling atoms to photonic crystal waveguides (PCWs) strong, efficient, and tuneable interactions mediated by light can be used for quantum simulations \cite{Cirac2012, hung2016quantum, chang2018colloquium}. Other opportunities include the coupling of motion or phonons with single atoms and photons to explore complex many-body physics \cite{NeumeierNorthupSingleAtomCavity}, to generate hyper-entanglement \cite{shaw25}, and to realize new nano-scale sensors operating in the quantum regime \cite{PhysRevLett.132.190001}.

Beyond fundamental research, the inherent integrated functionality of nanophotonic waveguides enables fast, efficient quantum networks interfaced with configurable atom-based nodes on chips \cite{kimble2008quantum}. The unique advantages of highly dissimilar physical systems (e.g., atoms, solid-state defects, quantum dots) could be combined in one and the same setting to realize efficient heterogeneous quantum memories, photon sources, and interaction zones controlled via stop-band modulators \cite{beguinpnas2020}. 

1-D atomic ensembles trapped in the evanescent field of a waveguide have been demonstrated \cite{vetsch2010optical, BackscatteringOff1DAtomString,BraggReflection1DChain}. There, the guided trapping light modes constrain the geometry of the trapped ensemble.
 
The deterministic and programmable coupling of ultracold atomic lattices with a designed pattern to 1-D and 2-D PCWs is difficult and requires significant multidisciplinary efforts in nanophotonics, atomic physics, quantum optics, and condensed matter physics. In \cite{beguin20, luan2020}, we proposed an experimental platform and designs of PCWs that solve significant challenges to combine nanophotonics and cold atom research.

Here, we demonstrate deterministic coupling of an ultracold atomic lattice to weak probe light propagating in suspended on-chip photonic waveguide circuits. A 2-D lattice of Cesium (Cs) atoms is prepared at a distance from a nanophotonic chip by a set of optical tweezer traps. The tweezer lattice is then moved into the evanescent field of a chosen 1-D waveguide. After a controllable time, atoms can be moved away from the waveguide and measured. The whole waveguide features photonic crystal sections with engineered dispersion properties for efficient interactions with Cs atoms. 

Our result represents a major experimental step toward realizing a platform that combines the capabilities of cold neutral atoms with the functionality and scalability of integrated nanophotonics within a single experimental setting \cite{chang2018colloquium, Gonzalez-Tudela2024}. Furthermore, we present new opportunities for in-situ sub-wavelength-resolution imaging of evanescent light and nanoscale dielectric structures using atoms as point-like detectors.
\begin{figure*}[t]
\centering
\includegraphics[width=1.0\linewidth]{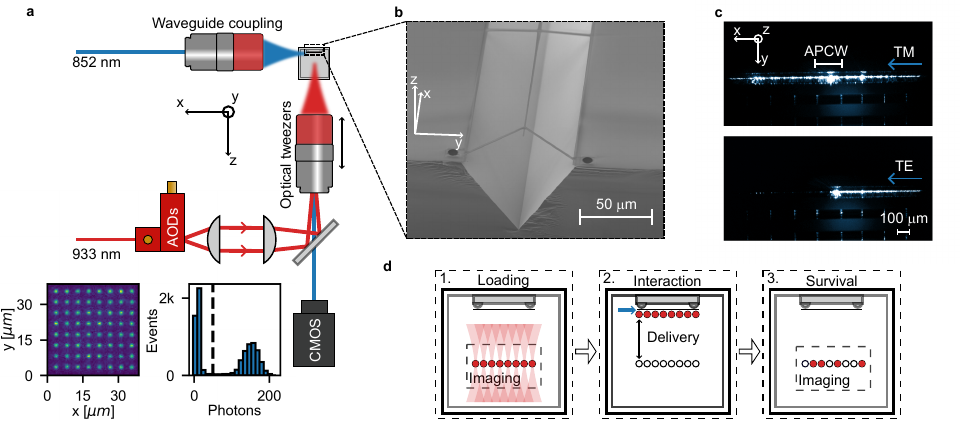}

\caption{\textbf{Details of the experimental setup.} \textbf{a} Overview of the experimental apparatus. A 2D array of atoms is trapped in optical tweezers at \SI{933}{\nano\meter} produced by a pair of crossed AODs. The tweezers are focused using a microscope objective mounted on a motorized stage allowing the atomic lattice to be brought from the loading region (\SI{3.6}{\milli\meter} from the chip) to the nanophotonic device. The atoms are imaged with a CMOS camera. The inset shows an example image of the 8x8 atomic array with a histogram of photon statistics. \textbf{b} Scanning electron microscope (SEM) image of a SiN device similar to the one used for this experiment. Our waveguides are suspended over V-grooves etched into the silicon substrate. \textbf{c}  Scattered light from impurities in the waveguide used in this work. Due to the TE bandgap of the APCW we can see a dependence of the waveguide transmission on the input polarization. Top: TM input polarization is transmitted through APCW. Bottom: TE input polarization is reflected by APCW. \textbf{d} Schematic of the experimental sequence. 1. The atoms are initially imaged in the loading region and are subsequently moved to the device. 2. A pulse of light on the D2 line is sent through the waveguide to interact with the atoms. 3. Finally, the atoms are imaged again after retracting the objective to evaluate the interaction.}
\label{fig:setup}
\vspace{-10pt}
\end{figure*}
\section{Results}
\subsection*{Combining photonics with cold atoms}
We realize a versatile light–matter interface by combining an array of single Cs atoms with light propagating in suspended nanophotonic waveguides. The atoms are held in tightly focused, near magic wavelength \cite{MagicWavelength} optical tweezers at \SI{933}{\nano\meter}, generated by a long-working distance objective lens with a nominal numerical aperture of NA = 0.42 (Fig.~\ref{fig:setup}a). Individual trap sites are dynamically reconfigurable using crossed acousto-optic deflectors, enabling agile construction of atomic arrays. The array is initially loaded (filling fraction $\sim50\%$) from an optical molasses \SI{3.6}{\milli\meter} away from the plane of the nanophotonic devices and is subsequently delivered into their close proximity with nanometer-scale precision by moving the tweezer lens with a closed-loop translation stage. This capability allows us to deterministically position atoms relative to the guided mode volume of the waveguides.  
An illustration of an experimental sequence is shown in Fig.~\ref{fig:setup}d. The atoms are initially loaded in the traps and imaged. They are then delivered to the nanophotonic device, where an interaction with the guided light occurs. Finally, the occupation of the remaining trapping sites is determined by moving the lattice back to the loading region and performing fluorescence imaging.

The nanophotonic devices, Si$_3$N$_4$ alligator photonic crystal waveguides (APCWs) \cite{gobanncomm}, exhibit strong light confinement and a polarization-dependent photonic bandgap (Figs.~\ref{fig:setup}b,c) \cite{luan2020}. These devices are suspended over V-grooves etched into the silicon substrate of the chip, reducing the impact of reflected trap light on the atoms when brought to the plane of the device. Free-space coupling of light from outside the vacuum chamber into the waveguide (Fig.~\ref{fig:setup}a) achieves estimated mode-overlap efficiencies for the transverse electric (TE, here $y$-polarized) mode of approximately 50\%, with the ability to further improve it to 90\% \cite{luan2020}. As shown in Fig.~\ref{fig:setup}c, the transmission through the APCW strongly depends on the input polarization, showing a dramatic reduction for the TE mode compared to transverse magnetic (TM).

\begin{figure*}[t] 
\centering
\includegraphics[width=1.0\linewidth]{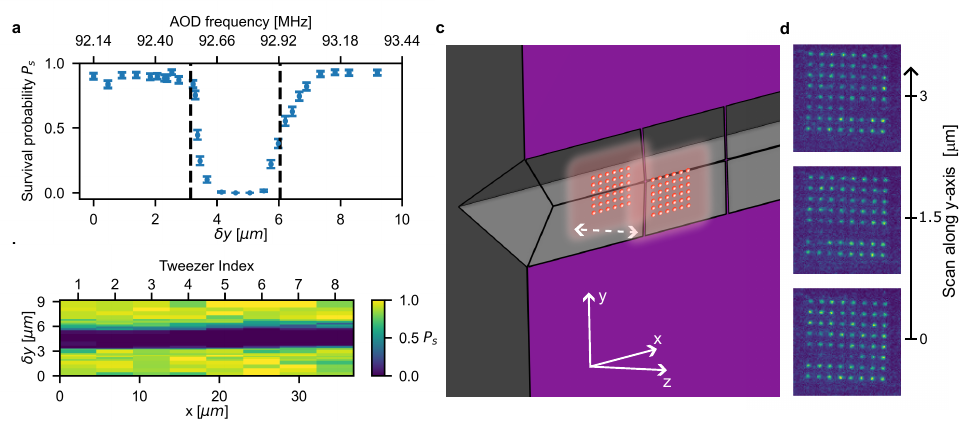}
\caption{\textbf{Scanning atom microscope}. \textbf{a} Plot of the survival probability ($P_s$) of a line of eight atoms as a function of the displacement ($\delta y$) with no light sent through the waveguide. Vertical dashed lines indicate the optical tweezer waist ($1/\mathrm{e}^2$ intensity radius  of \SI{1.2}{\micro \meter}). \textbf{b} Color map showing the survival probability of each atom in the array as a function of $\delta y$. Note the slight tilt of the waveguide with respect to the atomic array. \textbf{c} Schematic of the atomic array motion relative to the waveguide. The dashed line represents the trajectory the atoms follow when moved to the waveguide and back. \textbf{d} Images of the atomic array scanned across the y-axis after being brought in the proximity of the waveguide. Localized losses due to the presence of the waveguide are clearly observed. Furthermore, we can see a tilt similar to that shown in panel \textbf{b}.}
\label{fig:SAM}
\end{figure*}
\begin{figure*}[t]
\centering
\includegraphics[width=1.0\linewidth]{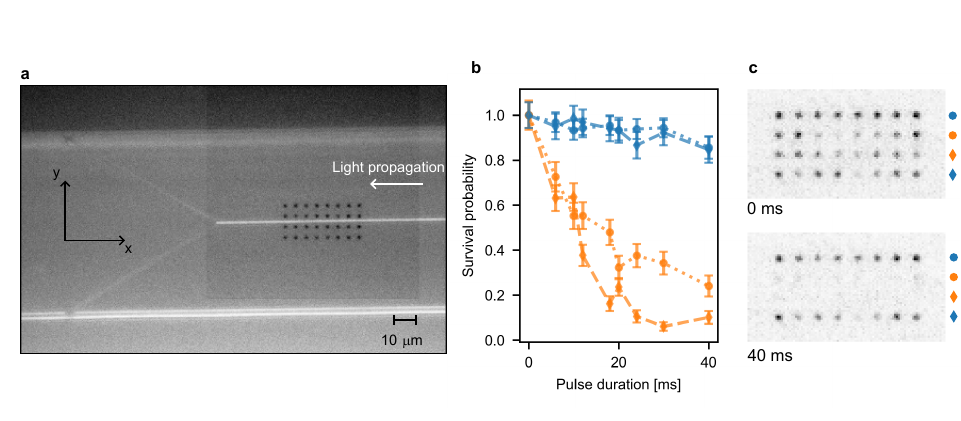}
\caption{\textbf{Atomic survival probability as a function of light pulse duration.} \textbf{a}. Image of the atomic array with a white-light image of the device superimposed. The white arrow indicates the direction of propagation of light through the waveguide. \textbf{b}. Survival probability for the inner (orange) and outer (blue) rows in the array as a function of the duration of the pulse of D2 light sent through the waveguide, for a constant light power (see main text). Data series are normalized to the zero pulse duration survival probability. Symbols differentiate the individual rows, see panel \textbf{c}. Dashed (dotted) lines serve as a visual aid. \textbf{c}. Averaged images of the atomic array for a \SI{0}{\milli\second} pulse duration (top) and a \SI{40}{\milli\second} pulse duration (bottom). The images illustrate the spatial dependence of the losses.}
\label{fig:rows_survival_PD}
\end{figure*}

\subsection*{Deterministic delivery}
We move the foci of the tweezers which trap the atoms using a smoothed trapezoidal motion profile along the longitudinal axis ($z$), enabling us to consistently transport the atoms to and from the waveguides with a survival probability exceeding 90\%. Distortions of the tweezer traps from surface reflection and cross-talks between multiple traps can reduce the overall delivery efficiency as compared to transporting a single tweezer-trapped atom.
The combined control of the tweezer array and the translation stage enables a three-dimensional embedding of atoms into the near-field of the waveguides. Fig.~\ref{fig:SAM}c schematically illustrates positioning of the atomic lattice  near the photonic waveguide by moving the tweezers focal plane. Here, the atoms are brought near the tapered edge-coupler of the waveguide, where the width is approximately \SI{180}{\nano\meter}. With the above capabilities, we perform systematic position scans for different array geometries, for instance, a row of eight atoms in the plane of the device or a square array as shown in Fig.~\ref{fig:SAM}d. The scanned lattice geometry can be adapted to the shape of the suspended dielectric structures. Scanning the position of a row of eight atoms, we observe that they remain stably trapped until the tweezers overlap with the device itself, at which point the survival probability drops sharply (Fig.~\ref{fig:SAM}a). The width of this lossy region is set primarily by the optical tweezer waist as indicated by the vertical dashed lines. The asymmetry in the survival between the two sides of the waveguide can likely be attributed to a slight misalignment of the tweezer-array plane w.r.t. the plane of the device. A relative tilt of approximately \SI{0.5}{\degree} within the xy-plane between the array and the waveguide is revealed when plotting the individual survival probability of the atoms (Fig.~\ref{fig:SAM}b). These measurements show both the possibility of using atoms to map out nanoscopic geometries and the ability to embed waveguides into the atomic array. The impact on an $8\times8$ array of atoms is shown in Fig.~\ref{fig:SAM}d. Here, the localized loss of atoms is readily apparent as the position is scanned across the waveguide. We have also been able to detect other features of the device, such as the angled tether arrays, using our atomic microscope.
These measurements provide a means to quantify both the geometry of sub-wavelength scale dielectric structures and tweezer-traps as well as the relative position alignment of individual atoms to the surface.

\begin{figure*}[t]
\centering
\includegraphics[width=1.0\linewidth]{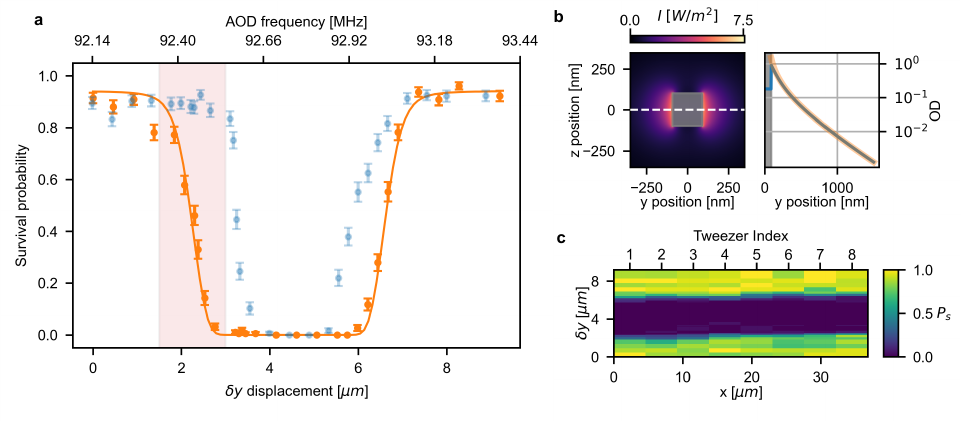}
\caption{\textbf{Measurement of the evanescent field with atomic arrays.} \textbf{a}. Atomic survival probability as a function of the displacement (AOD frequency) across the waveguide with (orange) and without (blue) light. Within the range of distances shaded in red we observe atomic loss solely due to interaction with the evanescent field, while no atomic loss is detectable at those distance without light in the waveguide. We are able to distinguish differences in the evanescent field with spatial resolution in the order of \SI{30}{\nano\meter}. The solid line shows a model prediction for the atomic survival probability using an asymptotic decay length of \SI{743}{\nano\meter}, inital temperature of $T=\SI{40}{\micro k}$ and optical power $P=\SI{400}{\pico\watt}$. Error-bars are given by the standard error on the mean. \textbf{b}. 
Intensity distribution of the fundamental TE mode in a $\SI{180}{\nano\meter}\times\SI{200}{\nano\meter}$ waveguide, obtained from MPB simulations. Gray areas indicate the waveguide. Left panel: The intensity distribution of the evanescent field in the $yz$ plane of the device. Right panel: Single atom OD along the direction of polarization of the evanescent field (indicated by the dashed line in the left panel) as a function of the y-position (blue) overlapped with a fit to equation~\eqref{eq:Evanescent Decay} (orange). \textbf{c}. Single-atom survival probability for the atomic array when light is sent through the waveguide.}
\label{fig:modes}
\end{figure*}
\subsection*{Near field light-atom interaction}
To probe the interaction between guided photons and the atomic array, we measure the atoms' survival probability as a function of the duration of a heating pulse, resonant with the Cs $6\mathrm{S}_{1/2} \mathrm{F=4}\to 6\mathrm{P}_{3/2} \mathrm{F'=5} $ transition. In Fig.~\ref{fig:rows_survival_PD}a we show a $4\times8$ array of atoms, placed symmetrically around the waveguide, such that only the central rows see an appreciable light intensity from the evanescent field. The optical power through the waveguide is approximately \SI{1}{\nano\watt}. The direction of light propagation through the waveguide is chosen such that the effect of uncoupled light on the atoms is minimized. This is achieved by sending the light from the far end of the waveguide, so that the intensity of the uncoupled light is significantly lower than that of the evanescent wave close to the waveguide. As shown in Figs.~\ref{fig:rows_survival_PD}b,c, only atoms in the two rows adjacent to the waveguide exhibit appreciable loss with increasing exposure time, whereas atoms located further away remain largely unaffected. Note the correlated behavior of the survival probability of the two central rows, owing to the strong position dependence of the field intensity.

We compare the position-dependent survival probability with and without guided light (Fig.~\ref{fig:modes}a). For this measurement, a \SI{6}{\milli\second} pulse with an estimated power of \SI{1}{\nano\watt} is sent through the waveguide. The displacement of the atoms in the plane of the device, $\delta y$, is scanned using one of the AODs. With guided light, we observe a strong increase in atomic loss in the vicinity of the waveguide due to the evanescent field of the guided mode. The shaded region of Fig.~\ref{fig:modes}a indicates a range of distances at which there is a pronounced effect of the light, without any additional loss due to the bare waveguide. 
The spatial dependence of the loss is further illustrated in Fig.~\ref{fig:modes}c, where the individual survival rate for the trapped atoms is plotted. Similarly to Fig.~\ref{fig:SAM}b, there is a detectable tilt between the waveguide and the atomic array, however the range of the loss is noticeably larger with light through the waveguide. These results show deterministic atom-photon coupling in our hybrid nanophotonic-cold atom platform. 

Numerical (MPB \cite{MPB}) calculations of the spatial dependence of the evanescent field for the TE mode of a \SI{180}{\nano\meter} wide waveguide are plotted in Fig.~\ref{fig:modes}b. The left panel shows the field distribution in the plane orthogonal to the propagation direction. Here, a clear anisotropy of the field is apparent. The right panel displays the calculated single-atom optical depth, $\mathrm{OD} = \frac{\sigma_0 I}{P}$, where $I$ is the local intensity, $P$ is the optical power, and $\sigma_0$ is the effective atomic resonant scattering cross-section for $\pi$-polarized light. This is plotted as a function of the distance from the waveguide center, taken in the device plane indicated by the dashed line in the right panel of Fig.~\ref{fig:modes}b. The mode decays rapidly close to the waveguide surface, slowing down at larger distances. The spatial dependence of the light intensity outside the waveguide can be approximated by the function 
\begin{equation}\label{eq:Evanescent Decay}
    I(r,\theta=0)=\frac{P}{\pi\rho}\frac{1}{r}\mathrm{e}^{-2r/\rho},
\end{equation}
where $P$ is the optical power and $\rho$ is the decay length of the evanescent field. A fit with this function is shown in orange in the left panel of Fig.~\ref{fig:modes}b, yielding a decay length of \SI{743.2\pm0.1}{\nano\meter}.

We estimate the survival probability as a function of position by modeling the motional state of the atoms with a truncated harmonic oscillator. Due to the finite Lamb-Dicke parameter in our traps, $\eta = \sqrt{\omega_r/\omega_T} \sim 0.26$, when driving the internal state of the atoms, there is a finite probability of changing the motional state, given by the Franck-Condon factors \cite{Wineland_2003} (see methods). Starting from a thermal state, the survival probability is calculated as the total population in the truncated harmonic oscillator Hilbert space after multiple scattering events, given by the scattering rate $R_\mathrm{sc}=\frac{\Gamma}{2}\frac{s}{s+1}$, where $s=I/I_\mathrm{sat}$ is the position-dependent saturation parameter, with $I_\mathrm{sat}=\hbar\omega_0 \Gamma/2\sigma_0$ for the linewidth $\Gamma$ and angular transition frequency $\omega_0$ of the Cesium D2 line. The solid line in Fig.~\ref{fig:modes}a shows a model prediction for the survival probability with realistic values for the model parameters, using a decay length of the mode of \SI{743}{\nano\meter}, an optical power of \SI{400}{\pico\watt} and an initial temperature of \SI{40}{\micro\kelvin}, consistent with release-recapture measurements after transport to and from the chip. The predicted survival shows good agreement with the experimental data and indicates that we understand the dynamics of the system quite well. These methods can be applied to different geometries to measure the mode profile of the evanescent wave with great precision.

Together, these results demonstrate the capability to realize a versatile quantum interface, combining single atoms and guided light through nanophotonic structures.  

\section{Discussion}
We have demonstrated the deterministic coupling of an ultracold atomic lattice to guided light in suspended on-chip photonic circuits. This achievement is based on several experimental advances and paves the way for new atom-light interfaces in the strong coupling regime, validating the nanophotonic–cold atom platform proposed in \cite{beguin20, luan2020}.

The use of high-stress Si$_3$N$_4$ allows us to combine suitable optical and mechanical properties with mature nanofabrication capability, enabling a scalable integration of complex photonic structures with atomic systems. Advances in coupler design have significantly improved the efficiency of light injection into the chip, with an overall transmission loss below \SI{3}{\deci\bel} as in \cite{beguinpnas2020}. However, the interaction with residual uncoupled light can mask the desired coupling to the guided modes. An effective design strategy to mitigate this limitation is to include waveguides with integrated bends, which can suppress stray light while preserving efficient coupling. 

We have also demonstrated the deterministic transport and delivery of single atoms and atom arrays over millimeter-scale distances with high retention rates, for which we provided a quantitative study. This capability establishes a bridge between free-space atomic arrays and integrated nanophotonic circuits. The finite confinement of the atoms in the traps means that the localization of the atomic motion is limited. This confinement can readily be improved by using higher NA objective lenses for atom trapping as well as cooling the atoms to the motional ground state after transport to the device.

The ability to position single atoms and atomic lattices within sub-micron to 100-nm distances from nanophotonic structures enables a variety of sensing modalities by exploiting both the internal and motional degrees of freedom of trapped atoms, for example, quantum electrometry \cite{pennetta2025}, the measurement of surface forces and vibrations \cite{balland2024}, and the detection of magnetic field gradients. We have shown that the spatial dependence of the evanescent field of guided light can be mapped out with a high level of precision. In-situ characterization of evanescent optical fields, particularly in photonic structures without high-Q resonators, is challenging. Here, we show that single atoms can serve as natural dipole probes of the local field, providing sensitivity down to the single-photon level or, alternatively, to very low intensities in the far field through long interrogation times. For example, the design of tapered waveguides or edge couplers can be studied by moving the atom lattice in the evanescent light along the waveguide axis. Unlike conventional techniques such as atomic force microscopy (AFM) or scanning near-field optical microscopy (SNOM), single atoms or atom arrays avoid the risk of damaging the sample by crashing a macroscopic probe tip into the surface, and enable three-dimensional probing of structures ranging from millimeter to nanometer scales. The shape of the atomic arrays or the “atomic tip” can be programmed and adjusted to the target geometries, benefiting from the quantum functionality of both individual atoms and collective states of atomic ensembles.  

The ability to transport atoms into the plane of the chip and beyond enables coupling to the fundamental TE mode of the dielectric nanostructures. The lowest-energy dielectric TE band of the 1-D photonic crystal waveguides gives rise to wide photonic band-gaps and atom-light interactions in the strong coupling regime per atom \cite{Yu14, hood2016atom, Burgers19}. For this platform, TE modes have lower transmission losses at the input couplers and tethers as compared to TM modes, as well as a larger mode overlap efficiency with a free-space gaussian mode \cite{luan2020}. In the present work, the guided light interacting with the atoms propagated first through the photonic crystal waveguide section, which can, in turn, be efficiently modulated via phonons \cite{beguinpnas2020}.

As we translate the atomic lattices further along the waveguide axis, we also observe the coupling of atoms to the more confined evanescent field of \SI{500}{\nano\meter}-wide waveguide sections. Further improvements are required to consistently bring atoms to regions of the waveguides with higher field confinement. This could be achieved by stabilizing the tweezer focal plane relative to the waveguides using an auxiliary beam. The suspended geometry also enables background-free imaging of atoms \cite{bernien2024} in close proximity to the waveguides.  A natural next step is to increase the coupling strength per atom and to interface atoms with new types of guided modes, such as those supported by photonic crystal waveguides. 

A necessary tool to harness light-mediated atom-atom interactions for quantum simulation is to map out the two-point Green’s function in quantum dielectrics. This can be achieved by using a single atom as a source of light and another single atom as an absorber or scatterer. Thus, combined with advancement in the deterministic preparation of single-atom arrays with large filling factor \cite{Grunzweig2010,barredo2016}, the probability to deliver at least two atoms per experimental cycle in the vicinity of nanophotonic waveguides can approach unity.

Beyond simple geometries, our results demonstrate that combining atoms with more complex nanophotonic device architectures, e.g., junctions, multiple parallel suspended nanobeams, and integrated cavities for optical tweezer traps, is within experimental reach.
This allows for future study of enhanced tweezers and new photonic geometries for hybrid on-chip traps \cite{beguin2020bpnas,benat2024,chenlung2024}.

Our platform provides a resource for a quantum interface between ultracold atomic arrays and light which can be used for generation of nonclassical spin states by quantum nondemolition measurement \cite{beguin2014generation,McConnell2015,Hosten2016,beguinprx2018} and for teleportation protocols connecting atomic arrays with other quantum systems.

\section{Methods}

\subsection{Platform description}
Atoms are sourced from Cs dispensers mounted in a dedicated source cell, approximately \SI{40}{cm} above the science cell containing the nanophotonic devices. A magneto-optical trap is used to capture a cloud of atoms. After polarization-gradient cooling, the atoms are dropped into a hollow gaussian beam (doughnut beam) and are transported to the science cell. Upon arrival, the atoms are slowed and cooled in an optical molasses. The array of optical tweezers is overlapped with the optical molasses and atoms are loaded into the traps, achieving a typical loading rate of $\sim50\%$. The trapping light is produced by a Spectra Physics Matisse Ti-Sapphire laser at \SI{933}{\nano\meter}. The fluorescence from the atoms is imaged on a Hamamatsu Orca Quest qCMOS camera. 

The array of optical tweezers is produced by a pair of acousto-optic deflectors (AODs), supplied by $2\times8$ direct-digital synthesizers. The use of AODs to generate the array, as opposed to spatial light modulators (SLMs) or digital-mirror devices (DMDs), mitigates the effects of reflection of trap light near the chip surface by shifting the beat frequencies of the traps well above the trap frequency. The focal plane of the optical tweezers is translated using a Physik Instrumente C-413 linear stage.

\subsection{Model for the light-matter interactions}
We model the survival probability of the atoms due to scattering of photons by considering the optical tweezer traps as truncated harmonic oscillator potentials. We experimentally determine the trap depth of the tweezers to be approximately \SI{340\pm11}{\micro\kelvin} with a radial angular trap frequency $\omega_T = 2\pi\times\SI{30.1\pm.1}{\kilo \hertz}$. This results in roughly 240 motional states of the harmonic oscillator potential. For each scattering event, the probability to transition from one motional state $m$ to another state $n$ is given by the Franck-Condon factors, 
\begin{align}
|\bra{n}D(\eta)\ket{m}|^2 & =  \mathrm{e}^{-|{\eta}|^2}\frac{m!}{n!}|\eta|^{2(m+n)}\,\,\times\nonumber \\
  & \bigg\lvert\sum_{k=0}^n \binom{n}{k}\frac{(-1)^k}{(m-k)!}|\eta|^{-2k}\bigg\lvert^2
\label{eq:fc}
\end{align}

where $\eta=\sqrt{\frac{\omega_r}{\omega_T}}$ is the Lamb-Dicke parameter and $\omega_r$ is the free-space angular recoil frequency for the Cesium D2 transition. To evaluate the survival probability for a given intensity and pulse duration, we start from a thermal state of the motion. For each scattering event, we apply eq.~\eqref{eq:fc} to the state vector. The survival probability is then given by the total population in the resulting state vector. Due to limitations in the size of the computed state space, only a maximum of 130 states were used. This leads to a slight reduction in the necessary power to eject an atom, as evidenced by the somewhat lower power used in the prediction in Fig.~\ref{fig:modes}a compared to experimental estimates. 

\subsection{Device description}
Details of our devices can be found in Refs.~\cite{beguin20,luan2020} and the references therein. The devices consist of adiabatically tapered edge-couplers, used to match the mode of a \SI{500}{\nano\meter} wide rectangular waveguide to a free-space Gaussian mode. The width of the edge-coupler for the device used in this work is \SI{180}{\nano\meter}. After the waveguide section, the waveguide subsequently splits into two parallel corrugated nanobeams, forming the alligator photonic crystal section. Coupling of light into the waveguides is achieved using a long working distance objective outside the vacuum chamber. The mode overlap efficiency is evaluated by coupling light originating from the edge coupler mode into an optical fiber. 

\section*{Acknowledgements}
We dedicate this work to the memory of Prof. H. Jeff Kimble and thank our former colleagues from his group at Caltech for their contributions to this emerging research field. We are grateful to Anders Simonsen for his work on the nanofabrication of our nanophotonic devices and their optical characterization during his time at the Niels Bohr Institute. We also thank our colleagues at HY-Q and QMATH for their support and the many fruitful discussions during the early stages of this project. This research was supported by the Novo Nordisk Foundation (Grant No. NNF20OC0059939 ‘Quantum for Life’ and Grant No. NNF24SA0088433 ‘Center for Biomedical Quantum Sensing‘) and by Villum Investigator Grant No. 25880.

\section*{Author Contributions}
JBB and ESP conceived the project. JTH conducted experimental work and theoretical analysis with the help from FG, JHM and JBB. JBM contributed with the fabrication and characterization of waveguides. JBB supervised the work assisted by JHM and ESP. 
All authors discussed the results and contributed to the final manuscript.



\bibliography{main}

\end{document}